\documentstyle[11pt,paspconf,epsf]{article}

\begin{document}

% added 1st page header for astro-ph distribution - 05/10/99
\vskip -1in
{\tt \noindent To appear in \underbar{Spectrophotometric Dating of Stars and
Galaxies},\\
eds.\ I.\ Hubeny, S.\ Heap, \& R.\ Cornett (ASP Conf.\ Series), 1999}

\title{Stellar Populations of Elliptical Galaxies from
Surface Brightness Fluctuations}
\author{Michael C. Liu}
%\author{Michael C. Liu\altaffilmark{1}}
\affil{Department of Astronomy, University of California, Berkeley, CA 94720}
%\altaffiltext{1}{\tt mliu@astro.berkeley.edu} 

\author{Stephane Charlot}
\affil{Institut d'Astrophysique de Paris, CNRS, 75014 Paris,
France}

\author{James R. Graham}
\affil{Department of Astronomy, University of California, Berkeley, CA 94720}

\begin{abstract}

We are using optical/IR surface brightness fluctuations (SBFs) to
validate the latest stellar population synthesis models and to
understand the stellar populations of ellipticals. Integrated light and
spectra measure only the first moment of the stellar luminosity function
($\Sigma n_i L_i$). Since SBFs also depend on the second moment ($\Sigma
n_i L_i^2$), they provide novel information, in particular about the
reddest, most luminous RGB and AGB stars, which are the most difficult
stars to model.  SBFs can also provide useful new constraints on the
age/metallicity of unresolved stellar populations in
ellipticals. Finally, developing accurate stellar population models
benefits several aspects of SBF distance measurements to galaxies.

\end{abstract}


\section{Introduction}

When observing a nearby elliptical or bulge of a spiral galaxy, there
are two distinct characteristics in the galaxy's surface brightness.
The first is the most obvious: the galaxy is brightest in the center and
grows fainter with increasing radius.  The second characteristic is only
apparent in good seeing conditions: the surface brightness is clumpy on
the scale of the seeing disk.  These clumps, which can be a few percent
of the mean surface brightness for the nearest galaxies like M~31 and
M~32, arise from Poisson statistical fluctuations in the number of stars
per seeing disk. Historically, this effect was known as ``incipient
resolution.''  In the modern context, they are called surface brightness
fluctuations (SBFs).

Tonry \& Schneider (1988) devised a technique to quantify SBFs and to
use them as a distance indicator for undisturbed early-type galaxies.
Specifically, they proposed using the ratio of the 2$^{\rm nd}$ moment
of the stellar luminosity function (LF) to the 1$^{\rm st}$ moment as a
standard candle:
\begin{equation}
%\bar{L} \equiv {{\Sigma\ n_i L_i^2}\over{\Sigma\ n_i L_i}}
\bar{L} \equiv \frac{\Sigma\ n_i L_i^2}{\Sigma\ n_i L_i}
\end{equation}
where $n_i$ is the number of stars of type $i$ with a luminosity of
$L_i$.  $\bar{L}$ has units of luminosity and is expressed in
astronomer's units as $\bar{M}$ and $\bar{m}$, the absolute and apparent
SBF magnitude, respectively.

\section{Utility for Stellar Population Studies}

Since $\bar{M}$ is an intrinsic property of the LF, SBFs can be useful
for studying the stellar populations of early-type galaxies.  The
potential parameter space to explore is very large, since the ``age''
and ``metallicity'' of the stars can be a complex combination of
formation epoch and subsequent history (e.g., Kaufmann, in this
volume). SBFs provide data {\em unique} from the integrated
light/spectra of these galaxies. Because of their $L^2$ dependence, SBFs
are very sensitive probes of the most luminous stars, the cool red giant
stars (Figure~1).
\begin{figure}
%\plotone{cumulative-Kband-color.eps}
%\plotfiddle{cumulative-Kband-color.eps}{2.5in}{0}{50}{50}{-180}{-10}
\plotfiddle{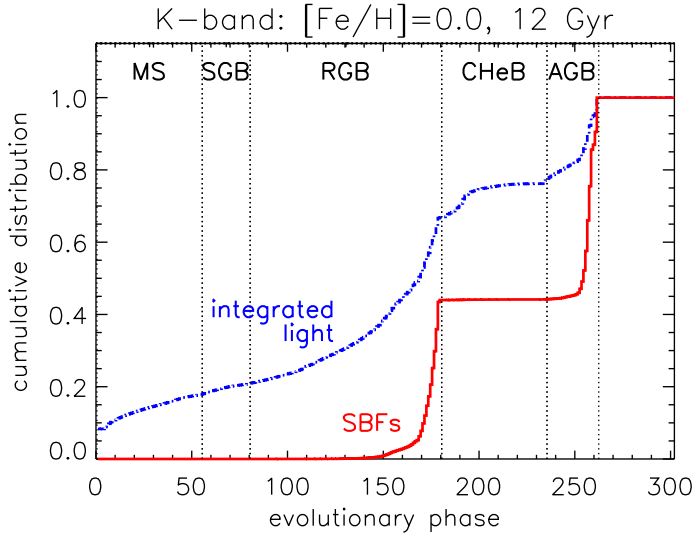}{2.5in}{0}{45}{45}{-180}{-15}
\caption{The cumulative contribution of different stars to the
\hbox{$K$-band} (2.2 \micron) integrated light and SBFs for a
single-burst population from Bruzual \& Charlot (1998). The x-axis goes
from the least evolved (main-sequence) stars on the left to the
most-evolved (AGB) stars on the right using an arbitrary index.  The
integrated light arises from stars of all phases; about 50\% comes from
the RGB, but the contribution of the different RGB phases is basically
degenerate. The SBFs originate only from the upper RGB and AGB, making
them a powerful probe of these stars in early-type galaxies.}
\label{fig-1}
\end{figure}
Modeling these stars' interior structure and emergent spectra is very
challenging.  Also, since bright RGB and AGB stars evolve quickly, only
a handful of each are present in any star cluster in the Milky Way or
Magellanic Clouds.  Therefore, SBF measurements, which arise from the
stellar population of entire galaxies, provide one of the best
observational tests of our current understanding of these stars.

\section{Models versus SBF Data}

Recent observational and theoretical advances make now a ripe time to
revisit and expand SBF stellar population studies.  Past modeling (Tonry
et al.\ 1990; Buzzoni 1993; Worthey 1993) used the previous
generation of evolutionary tracks (VandenBerg or Revised Yale
Isochrones) based on older stellar opacities.  The optical SBF dataset
has been expanded and improved considerably (Tonry et al.\ 1997), and a
growing body of IR data has widened the spectral coverage.

Figure~2 presents $K$-band SBF data for Fornax cluster ellipticals from
Liu et al.\ (1999b) and Jensen et al.\ (1998).
\begin{figure}
%\plotone{kbar-semiemp.ps}
%\plotfiddle{kbar-semiemp.ps}{2.5in}{90}{50}{50}{180}{-50}
\plotfiddle{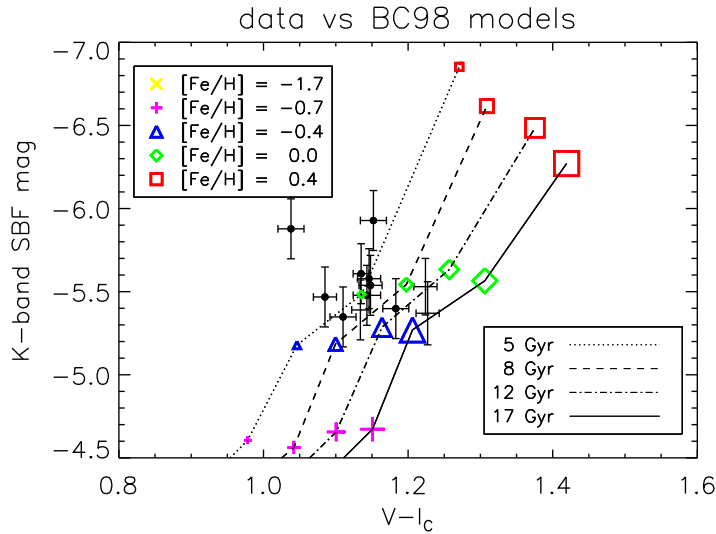}{2.5in}{90}{45}{45}{150}{-50}
\caption{$K$-band SBF data for Fornax cluster ellipticals versus BC98
models.  Models with the same [Fe/H] have the same symbol, and lines
connect models of the same age. The SBF magnitude is expected to be very
metallicity-dependent so the data imply the galaxies have comparable
metallicities but a spread in ages.}
\end{figure}
Fornax is appealing since it is close enough to have a Cepheid distance
measured with HST (Silbermann et al.\ 1999) and is compact on the sky,
implying the galaxies basically lie at the same distance.  Predictions
from Bruzual \& Charlot (1998) single-burst models are overplotted;
these use Padova evolutionary tracks, stellar spectra of Lejeune et al.\
(1997), and a semi-empirical AGB prescription (Charlot \& Bruzual
1991). (See Bruzual, in this volume.) The good agreement between data
and models means the lifetimes, luminosities, and colors of the bright
RGB and AGB stars are roughly correct, or else multiple errors are
cancelling each other out (e.g., Charlot et al.\ 1996).

The BC98 models agree worse with the $I$-band empirical SBF calibration
(Tonry et al.\ 1997) than with the $K$-band data, which has lead us to
revise the most evolved stars in the models (Liu et al.\ 1999a).
%The empirical $I$-band SBF calibration (Tonry et al.\ 1997) and BC98
%models agree poorer than at $K$-band, which has lead us to revise the
%most evolved stars in the BC98 models (Liu et al.\ 1999b). 
This
illustrates the usefulness of SBF tests.  Although the models agree
reasonably well with integrated colors of elliptical galaxies and Local
Group globular clusters, it is the SBF comparison which reveals possible
weaknesses in the modeling of the most luminous stars.

\section{New Tools for Breaking the Age/Metallicity Degeneracy}

In old populations, $\bar{M}$ is expected to be very metal-dependent,
since it strongly tracks the RGB and AGB, whose colors are governed by
metallicity (Frogel et al.\ 1983).  Therefore, SBF data are potentially
very useful in characterizing the stellar content of ellipticals.
Broad-band colors are degenerate in age and metallicity, with
changes of $d({\rm log}\ age)/d({\rm log}\ Z)\approx3/2$ preserving the
color (Worthey 1994 [W94]; Worthey, this volume).  Absorption line
indices can be more age- or metal-sensitive, e.g., $H\beta$ and
$H\gamma$ are age-sensitive ($\la 1.0$) and Mg$_2$ and C$_2$4668 are
metal-sensitive ($\approx 2-5$).  In comparison, IR SBF magnitudes are
predicted by the BC98 and W94 models to have \hbox{$d({\rm log}\
age)/d({\rm log}\ Z)\ga6$} in old populations.  This suggests IR SBF
data combined with Balmer absorption indices could effectively
disentangle age and metallicity effects in ellipticals.

\section{SBF Distances: Building a Better Standard Candle} 

Developing accurate models is critical for measuring distances to
galaxies from SBFs.  Optical/IR SBF distances can be measured to
$cz\approx10^4$~km/s and therefore to obtain $H_0$, but accurate
distances rely on accurate $K$-corrections to account for the
redshifting of the galaxy spectrum.  Moreover, models can guide us to
better observations. For instance, $I$-band SBF distances use the
$(V-I_C)$ integrated galaxy color to account for stellar population
variations from galaxy to galaxy (Tonry et al.\ 1997). Both the BC98 and
W94 models suggest that $(V-K)$ or $(I-K)$ colors, which are more
metallicity-driven, would lead to a better correction, reducing the
scatter by $\ga$50\%. The wider spectral ranges of these colors also
lead to greater tolerance of errors in photometry or reddening than
$(V-I_C)$. Finally, accurate models could provide a purely theoretical
calibration for SBF distances, independent of any Cepheid calibration.

%\acknowledgments
%MCL thanks NOAO/CTIO for travel support.


\begin{references}
%\parskip=1pt
\baselineskip=2pt
\reference Buzzoni, A. 1993, \aap, 275, 433
\reference Bruzual, G. \& Charlot, S. 1998, in preparation (BC98)
\reference Charlot, S. \& Bruzual, C. 1991, \apj, 367, 126
\reference Charlot, S., Worthey, G., \& Bressan, A. 1996, \apj, 457, 625
\reference Frogel, J. A., Cohen, J. G., \& Persson, S. E. 1983, \apj,
275, 773
\reference Jensen, J. B., Tonry, J. L., \& Luppino, G. A.\ 1998, \apj,
505, 111
%\reference Kuntschner, H. \& Davies, J. 1998, \mnras, 295, L29
\reference Lejeune, Th., Cuisinier, F., \& Buser, R.\ 1997, \aaps, 125, 229
\reference Liu, M. C., Charlot, S., \& Graham, J. R. 1999a, in preparation
\reference Liu, M. C., Graham, J. R., \& Charlot, S. 1999b, in preparation
\reference Silbermann, N. A. et al.\ 1999, \apj, 515, 1
\reference Tonry, J. L. \& Schneider, P. 1988, \aj, 96, 807
\reference Tonry, J. L., Ajhar, E. A., \& Luppino, G.\ 1990, \aj, 100, 1416
\reference Tonry, J. L., Blakeslee, J. P., Ajhar, E. A., \& Dressler,
A.\ 1997, 475, 399
\reference Worthey, G. 1993, \apjl, 409, 530; erratum 1993, ApJ, 418, 947
\reference Worthey, G. 1994, \apjs, 95, 107 (W94)
\end{references}
\end{document}